**The standing pool of genomic structural variation in a natural population of *Mimulus guttatus***

**RUNNING TITLE**

**Indel variation in *Mimulus guttatus***

**AUTHORS**

Lex E. Flagel[1,2,3], John H. Willis[1], and Todd J. Vision[2]*

[1] Department of Biology, Duke University, Durham, NC, USA

[2] Department of Biology, University of North Carolina, Chapel Hill, NC, USA

[3] Present address: Monsanto Company, Chesterfield, MO, USA

* Correspondence to: tjv@bio.unc.edu

**ABSTRACT**

Major unresolved questions in evolutionary genetics include determining the contributions of different mutational sources to the total pool of genetic variation in a species, and understanding how these different forms of genetic variation interact with natural selection. Recent work has shown that structural variants (insertions, deletions, inversions and transpositions) are a major source of genetic variation, often out-numbering single nucleotide variants in terms of total bases affected. Despite the near ubiquity of structural variants, major questions about their interaction with natural selection remain. For example, how does the allele frequency spectrum of structural variants differ when compared to single nucleotide variants? How often do structural variants affect genes, and what are the consequences? To begin to address these questions, we have systematically identified and characterized a large set submicroscopic insertion and deletion (indel) variants (between 1 kb to 200 kb in length) among ten individuals from a single natural population of the plant species *Mimulus guttatus*. After extensive computational filtering, we focused on a set of 4,142 high-confidence indels that showed an experimental validation rate of 73%. All but one of these indels were < 200 kb. While the largest were generally at lower frequencies in the population, a surprising number of large indels are at intermediate frequencies. While indels overlapping with genes were much rarer than expected by chance, nearly 600 genes were affected by an indel. NBS-LRR defense response genes were the most enriched among the gene families affected. Most indels associated with genes were rare and appeared to be under purifying selection, though we do find four high-frequency derived insertion alleles that show signatures of recent positive selection.

**AUTHOR SUMMARY:**

The advent of inexpensive sequencing technologies has revealed an abundance of submicroscopic structural variants (in the kilobase to megabase size range) in a number of different eukaryotic model systems. Surveys of natural populations are needed to understand the prevalence of structural variants in the wild, the forces contributing to their maintenance, and the extent to which they contribute to genetic adaptation. Here we study over 4,000 novel large indel polymorphisms in a natural population of the plant species *Mimulus guttatus*. Plant defense response genes are significantly enriched among the nearly 600 genes structurally affected by large indels. Nonetheless, as a group, large indels are under-represented in genes and over-represented in transposable elements and other non-coding parts of the genome. Population genetic data suggest that the frequencies of most large indels are depressed by purifying selection. Four novel insertion alleles are associated with regions of the genome apparently under positive selection.

**INTRODUCTION**

From comparative genomics we know that a large portion of the genetic differences between closely related species are structural in nature rather than substitutions involving single nucleotides [1,2,3]. These structural variants (SVs) include insertions, deletions, inversions, and transpositions. Many SVs are now fixed within species, but must have arisen and increased in frequency for some period of time prior to reaching fixation. From this we might anticipate there to be a significant pool of standing genetic variation for SVs within species or even within populations. However, until recently, SVs could only be studied if they were large enough to have a visible manifestation, such as the inversions seen in *Drosophila* salivary chromosomes [4]. With the advent of new technologies, we now have an unprecedented ability to discover much smaller "submicroscopic" SVs in a systematic manner on a genome-wide scale. With these tools we can begin to address the contribution of SVs to the total pool of genetic variation in a population and determine how these polymorphisms interact with natural selection.

Various genomic technologies have allowed researchers to detect and catalog submicroscopic SVs [5]. The first wave of techniques involved microarrays specifically built to interrogate the genome with oligonucleotide probes [6]. Recent approaches have capitalized on next-generation sequencing as a vehicle to generate millions of paired-end reads, which can be compared to a reference genome to discover SVs [7]. These paired-end reads are small genomic fragments (~ 500 bp) sequenced incompletely from both ends, creating a single-stranded 5' to 3' read on each strand with 200-300 bp of unsequenced insert in between. When comparing a newly sequenced accession to a reference genome, deviations from the expected insert size and expected read pair configuration can be used to identify SVs. Paired-end sequencing offers considerable improvements in SV resolution compared to earlier microarray based technologies [7,8], and has the added benefit of also exposing single nucleotide polymorphisms (SNPs).

Studies of model organisms such as *Arabidopsis*, *Drosophila*, humans, and maize have revealed considerable levels of segregating structural polymorphism [9,10,11,12,13,14,15]. In the case of humans, the number of nucleotide differences between individuals due to SVs is reported to be greater than that due to SNPs [16]. While many of these studies have compared genotypes sampled from throughout the species' range, recent studies in humans, three-spined stickleback and *Drosophila melanogaster* [17,18,19,20] have shown that polymorphic SVs also contribute to the standing genetic variation within populations.

Here, we report a population genomic analysis of a large and representative sample of SVs segregating within a single well-studied natural population of the plant species *Mimulus guttatus*. The SVs were discovered through paired-end whole-genome shotgun sequencing of 10 inbred lines from the focal population. We also sequenced two inbred lines from distantly related populations in order to determine the derived and ancestral allele for each polymorphism. We restrict our focus to large indels because we find that they are the class of submicroscopic SV that could be most reliably validated *in silico*. In order to provide a glimpse into the role of selection in shaping the population genomic diversity of large indels, we explore their spatial distribution in the genome, compare their allele frequency spectrum to that of single nucleotide polymorphisms (SNPs), and test whether any large indels may be increasing in frequency due to direct or indirect or positive selection.

**RESULTS**

**Identifying SVs from Paired-End Sequence Data**

To identify candidate SVs, we obtained approximately 531 million paired-end sequences from 12 inbred accessions. Ten of these were derived from the focal population (Iron Mountain, Oregon) and two from outgroup populations (**Table 1**). Nine of the individuals from the focal population were chosen at random from a collection of approximately 250 viable inbreds extracted from the Iron Mountain population, while the tenth was the inbred accession used to create the reference genome.

The principal challenge in reliably detecting SVs from alignment of paired-end sequences against a reference is that chimeric read pairs and erroneous alignments generate a large number of false-positive SVs [7], and in fact many of the initial candidate SVs in our dataset are physically impossible to reconcile with one another (**Fig. S1**). Thus, it was essential to filter SV predictions by exploiting additional signals in the data.

We employed a progression of filtering steps, summarized below and described more fully in the **Materials and Methods.** The first filtering strategy was based on the data from resequencing the inbred accession used to create the reference genome. All abnormally aligned read pairs deriving from the reference accession, when aligned to itself, serve as a marker for regions of the genome that produce unreliable read alignments (**Fig. S1**).

Various classes of SV were further filtered based on expectations for patterns that would be seen in either false- or true-positives. For example, we removed putative indels that did not show the expected low read coverage within the putatively deleted interval. For inversions and transpositions, we removed any candidates that fell in genomic windows with elevated rates of abnormally aligned reads. These filters greatly reduced the number of putative SV calls under consideration. Of the initial set of 13,845 putative indels, 4,142 high-confidence indels were retained. For inversions, the reduction was from 141 putative events to 35. And for transpositions, the filtering steps reduced 716 putative events to just 3. The fate of all abnormally aligned reads and their attrition as a result of various filters is available as a supplementary data file at the Dryad digital repository [21].

**Validation by PCR assays**

We then sought to experimentally validate a subset of the candidates passing these filters using PCR assays that would yield different amplicons depending on which allele was present in the sample. For inversions and transpositions, many of the putative events derived from areas of the genome that consistently produced unreliable alignments. We attempted to validate seven of the retained inversions using PCR (those shown in **Fig. S2**), but were unable to develop unique primer pair combinations as all were located within regions rich in repetitive sequences. We did not attempt to validate any of the 3 transpositions. While there probably are true inversions and transpositions among the accessions we resequenced, it remains a challenge to reliably identify them using paired-end reads. Our experience mirrors that described by the 1000 Genomes Project [15], and, like these authors, here we focus our analyses on the high-confidence indels only.

We developed PCR assays to test the existence of 48 indel allele pairs. These assays were done using either one or two amplifications. For smaller indels we used one PCR amplification external to the predicted indel boundary. A genotype homozygous for a deletion allele was expected to produce a small amplicon of an expected size, while a genotype with the insertion allele was expected to produce a longer reference length amplicon. For longer indels it was not possible to amplify across the insertion allele so we used two amplifications, one with both primers external to the predicted indel boundaries, and one with one internal and one external primer. A genotype homozygous for a deletion allele is expected to produce an amplicon only with the external primer pair, while a genotype with an insertion allele is expected to produce a reference-length amplicon using only the internal/external primer combination. Among the 48 assayed indel predictions, 24 confirmed our *in silico* predictions, 9 showed evidence of a false-positive call, and 15 were inconclusive due to failed PCR amplification. Of the 33 conclusive assay results, the overall rate of validation was 72.7% (**Table S1**), a figure comparable to a recent study in humans [17]. From these results, we conclude that the majority of high-confidence indels are true positives.

We note that our dataset of indels does not include any insertions relative to the reference, since we only detect putative indels by having two read pairs aligned at positions separated by at least 1000 bp in the reference genome (see **Materials and Methods**).

**Population Genomics of Indel Alleles in the *Mimulus* Genome**

To determine the accuracy of the allele frequencies observed in the initial sample of 10 resequenced lines, we genotyped 14 of the validated indels in a larger number of inbred lines (average $N$ = 128) extracted from the focal population (Iron Mountain). Assuming that allele frequencies in this large sample approximate the true population frequencies, we can estimate the error of the initial frequency estimates. The Pearson's correlation coefficient between the frequency estimates was 0.83, and the slope of the relationship was 0.77. The average absolute difference between both estimates for all 14 indels was 0.126 (**Fig. S3**), which is only slightly higher than the expected value of 0.097 (see **Materials and Methods**). Because the slope is near 1 and the average absolute difference is near its expected minima when accounting for sampling error, we conclude that an estimate of allele frequency from the initial sample can be roughly used as a proxy for population frequency.

Next, we checked to see whether the variable frequency of indels among the nine non-reference lines could be a result of variable sequencing depth. The number of indels observed and sequencing depth show little association, and in fact the Pearson's correlation coefficient is slightly negative ($r$ = -0.175; **Fig. S4**). The number of indel differences from the reference genome is better explained by the overall patterns of nucleotide genetic differentiation among the lines, as evidenced by a positive correlation between SNP divergence from the reference genome and indel divergence from the reference genome (Pearson's $r$ = 0.582; **Fig. S4**). From this, we conclude that we have sequenced at sufficient depth to detect low to intermediate frequency indel alleles.

Indels in the Iron Mountain population ranged from 1,000 bp up to 204 kb with a median size of 2,562.5 bp (**Table 2**). We asked whether large indels were disproportionately rare, as might be expected if larger indels are subject to stronger purifying selection. Because the ten accessions we resequenced were highly inbred (**Table 1**), we observe only one allele per accession, and the minor allele frequency (MAF) ranges from 0.1 to 0.5 by increments of 0.1. Within the Iron Mountain population, the mean MAF for all 4,142 indel polymorphisms was 0.255 (**Table 2**). The mean indel size for each of the 5 MAF categories is as follows: 0.1 = 6,665 bp, 0.2 = 4,869 bp, 0.3 = 4,807 bp, 0.4 = 4,386 bp, and 0.5 = 4,307 bp. Indels with higher MAF tended to be smaller and have a more restricted upper size range when compared to indels with a low MAF (**Fig. 1**), a result that is statistically significant (Welch's one-way ANOVA; $F$ (4, 1,829.5) = 7.33, $P = 7.4 \times 10^{-6}$). This observation suggests that purifying selection tends to be stronger for large indels, removing them from the population before they reach an appreciable allele frequency. Nonetheless, there exist a pool of large indels segregating at intermediate frequencies. For example, 9% of the indels with a MAF of 0.3 or higher are greater than 10 kb.

To provide insight into how selective forces differ for large indels and SNPs we compared their population genetic and genomic distributions. We extracted approximately 1.3 million SNPs from the resequencing data to compare the frequency and distribution of SVs and SNPs. The SNPs were partitioned into coding and non-coding polymorphisms and the coding SNPs were further partitioned into synonymous and nonsynonymous polymorphisms. The average minor allele frequency was similar for all classes of polymorphism, ranging from 0.218 to 0.277 (**Table 2**). **Fig. 2** shows the cumulative allele frequency spectrum and Tajima's D values for indels and SNPs. All values shown in **Fig. 2** fall within the 95% confidence interval expected for polymorphisms experiencing neutral evolution as determined by coalescent simulations.

We used two distantly related individuals from outgroup populations (**Table 1**) to polarize indels and SNPs. The minor allele for a large indel is more likely to be derived than it is for a SNP (0.718 for indels compared to 0.619 for SNPs; $\chi^2$ test of independence $P = 4.2 \times 10^{-13}$). This pattern could arise because indels have a higher mutation rate than SNPs, because purifying selection purges indels from the population faster than SNPs, or some combination of the two factors. We note that the SNP and indel polymorphisms examined here likely are less deleterious when homozygous than a sample drawn from nature. This is because the studied accessions were inbred by selfing for several generations prior to sequencing (**Table 1**). By rapidly reducing heterozygosity, selfing exposed recessive mutations to atypically strong selection, likely purging many of the most deleterious alleles.

Across the *Mimulus* genome, there is considerable heterogeneity in the abundance of SNPs and large indels. To compare the distributions for the two types of polymorphism we partitioned the genome into non-overlapping 500 kb windows and tallied the segregating indels and SNPs in each window. To normalize for sequencing coverage difference between windows, we divided the indel and SNP tallies by the total read coverage among all lines for each window. We then fit a linear model between the normalized indel and SNP counts in each window. We find a slight – albeit significant – positive relationship between the abundance of indels and SNPs in the genome (Pearson's $r$ = 0.208; slope = 4.9×10$^{-4}$; $df$ = 662; $P$ = 6.3×10$^{-8}$). However, SNP density explains only about 4% of the variation in indel density. This suggests these mutational processes are weakly correlated throughout the genome, at least at the scale of 500 kb windows.

**Indels in Genes and Transposable Elements**

There are approximately 27,000 genes and 239,000 transposable elements (TEs) or TE fragments annotated in the *M. guttatus* genome, comprising 24.2% and 56.5% of the assembled nucleotides, respectively. We wanted to determine how indel polymorphisms were distributed with respect to these features. First, we isolated the alignable portions of the genome using nucleotide positions from the resequenced reference genome accession (IM62) with mapping qualities ≥ 29 as a guide (see **Materials and Methods**). Following this filtering step, the alignable fraction was composed of 39.9% genes and 35.1% TEs, reflecting the relative uniqueness of gene sequences and the concentration of unalignable repetitive sequences among TEs (**Fig. S5**). The proportions from the alignable fraction of the genome represent the expected null genomic distribution of polymorphisms in our data after accounting for the bias of sequence alignment.

Indels were assigned as genic if the indel interval intersected with an annotated gene, including both coding and non-coding (introns) gene components. The same was done for annotated TEs. We find a strong enrichment of indels in TEs and corresponding paucity of indels in genes. The density of indels in TEs is 244.8 indels per Mb of TE sequence. Collectively these account for 79.3% of all observed indel nucleotides, as compared to 35.1% expected under the null model. In contrast, there are 48.2 indels per Mb of genic sequence, accounting for only 3.5% of all observed indel nucleotides, as compared to 39.9% expected under the null model (**Fig. 3A**). At the nucleotide level, the nonrandom distribution among these genomic categories is highly significant ($P < 2.2 \times 10^{-16}$ in a $\chi^2$ test for independence). These results suggest that there is strong purifying selection against the majority of large indels that arise within genes.

By comparison, 27.4% of all observed SNPs are in coding regions (**Table 2**), nearly 8-fold higher than the corresponding proportion for large indels. The SNPs within coding sequences are predominantly synonymous (**Table 2**), but even nonsynonymous SNPs comprise a higher proportion of all SNPs (8.0%) than the proportion of indel nucleotides in genes (3.5%). Under the assumption that SNPs and large indels mutations occur at roughly the same frequency throughout the genome, this result would suggest that the average large indel in a gene experiences stronger purifying selection than the average nonsynonymous mutation. By contrast, large indels in TEs appear to be under weaker purifying selection than synonymous SNPs.

Despite the fact that only a small fraction of the observed indels occur within genes, we do find 414 indels disrupting 598 genes among the nine non-reference accessions. Comparing between the resequenced inbred lines, we find an average pairwise difference of 204 indel containing genes, or approximately 0.7% of all annotated genes. Assuming deletion alleles for these genes are predominantly nonfunctional and recessive, we would expect that they could be complemented in a hybrid containing a functional allele. That is, if we made synthetic hybrids between any of the non-reference Iron Mountain inbred lines sampled here we'd expect the average hybrid to mask 204 nonfunctional recessive genes contributed by its parents.

We were interested to see what types of genes are affected by indels. Using *M. guttatus* paralogous gene family clusters from Phytozome (www.phytozome.net; version 8.0) we found that genome-wide the median cluster includes six genes. For genes with a large indel the median cluster size is 21 genes. This difference is significant (Mann-Whitney U test $P < 2 \times 10^{-16}$), and provides evidence for enrichment of indels among large gene families. This finding is consistent with the hypothesis that selection is weaker against indels in large gene families with many redundant paralogs when compared to small gene families [22], but may also reflect an elevated indel mutation rate through illegitimate recombination among paralogs. The most affected gene families include a putative nucleotide-binding site leucine-rich repeat (NBS-LRR) family (Phytozome v8.0 gene family #31803493), two putative F-box domain families (Phytozome v8.0 gene family #31838851 and #31808429), and a putative cytochrome P450 family (Phytozome v8.0 gene family #31803960).

To understand where large indels tend to occur in genes, we determined their distribution across various genic components. The indels are slightly, albeit significantly, enriched in introns and underrepresented in UTRs and exons (all $\chi^2$ goodness-of-fit $P < 0.001$; **Fig. 3B**). We also looked for spatial patterns at finer resolution. To accomplish this, we first partitioned all indel-containing genes into three functional components, the 3' and 5' UTRs and the gene body (exons and introns), and then within each component normalized to a standard length. For both UTRs, there is significant enrichment of indels distal to the coding portion of the gene (**Fig. 3C**). Within the gene body there is significant enrichment on the 5' and 3' extremities. These results suggest that indels tend to accumulate near the periphery of all three genic components. A similar result has been seen for small indels (< 60 bp) in humans [17]. These patterns may reflect weaker selection against indels affecting peripheral components of genes when compared to more direct hits.

We then looked at the 1,855 (44.8%) large indels that are at least 90% constituted by a single annotated TE. These indels are likely associated with the mobility of a single element, and can be used to estimate the relative activity of various TE classes (**Table S2**). The most abundant class of TEs associated large indels are MULEs, which account for 464 indels (25% of all indels constituted by a single annotated TE). This is an enrichment of approximately 1.6-fold relative to the frequency of MULEs among annotated TEs in the reference genome (Fisher's Exact Test $P = 1 \times 10^{-8}$). On the other end of this spectrum are Gypsy and helitron elements. Each shows 1.4-fold reduction relative to expectation (Fisher's Exact Test $P = 3.8 \times 10^{-6}$ and $9.5 \times 10^{-7}$, respectively). Following Bonferroni correction, the families with significantly higher than expected activity are MULE and TRIM, while those families with significantly lower than expected activity are helitron, Gypsy, and LARD (**Table S2**). Among the eight most over-represented TE families, seven are class II "cut-and-paste" DNA transposons, while all but one of the five class I retrotransposon "copy-and-paste" families were represented at expected levels or significantly underrepresented. We note that our indel discovery strategy allows the detection of novel insertions alleles only if they include the reference line, or deletion alleles among any combinations of the other 9 non-reference lines. This creates a bias in favor of finding deletions relative to insertions, which in turns favors the discovery of polymorphic class II DNA transposons. This bias likely plays a role in differentiating the activity of class I from class II TEs. That said, there remain considerable differences among elements of the same class, and these contrasts should be unaffected by the discovery bias noted above.

**Strength of Selection on Large Indels**

Alleles under positive or negative selection are expected to be on average younger than neutral alleles found at the same frequency in a population [23]. Using a diffusion approximation that assumes no dominance and a constant population size, Maruyama [23] showed that the expected mean allele age is symmetric for positive and negative selection coefficients of the same magnitude. Kiezun *et al.* [24] extended this result to create an estimate of the strength of selection on different types of rare mutation using the relationship between intra-allelic nucleotide diversity ($\pi_A$) and mean allelic age. We apply this method to our data to estimate the strength of selection on the subset of large indels within genes relative to synonymous and nonsynonymous mutations. To do this we first partition the three mutation types based on their observed allele frequency in the sample population. Then, conditioning on alleles at the same frequency, we estimated $\pi_A$ in a 500 bp haplotype centered on the focal mutation. We chose this small haplotype size to minimize the probability that our intra-allelic sample has experienced a recombination event.

For all synonymous mutations across all allele frequencies the grand mean $\pi_A$ was 0.0087, which can be interpreted as the expected diversity accumulated over the average coalescent time for a neutral allele in the population (i.e. after $2N_e$ generations). Mean $\pi_A$ for genic deletion alleles, nonsynonymous mutations and synonymous mutations at 20% allele frequency are 0.0031, 0.0047, and 0.0055, respectively. All values are lower than the grand mean $\pi_A$, which is expected because conditioning on alleles at 20% frequency should bias the pool toward younger − and hence less diverse − intra-allelic haplotypes. By dividing by the grand mean $\pi_A$ for all synonymous mutations, mean $\pi_A$ for alleles at 20% frequency can be converted to average coalescent times – expressed in $2N_e$ generations – of 0.36, 0.54, and 0.63 for genic deletion alleles, nonsynonymous mutations, and synonymous mutations, respectively (**Fig. S6**). The lower values observed for genic deletions and nonsynonymous mutations imply younger alleles and stronger selection on average for these loci when compared to putatively neutral synonymous loci, and that the difference from neutrality is greater for genic deletions than for nonsynonymous mutations.

With these coalescent time estimates, we can approximate the magnitude of the average coefficient of selection on each mutational class using a diffusion approximation [24]. The magnitude of the mean population scaled coefficient of selection ($\bar{\gamma}$) on synonymous mutations at a 20% allele frequency is estimated to be approximately |1.9|, while nonsynonymous mutations are |2.7|, and genic deletions are |5.4| (**Fig. S6**). For alleles at a 20% frequency, genic deletions are the youngest class and are estimated to experience 2.8 times stronger selection than synonymous mutations. Neutral alleles are expected to be ≤ |1|, thus all loci at a 20% frequency are younger (i.e. harbor on average lower $\pi_A$) than would be expected under neutrality, even synonymous polymorphisms. Because the expected allele age is symmetric for positive and negative selection coefficients of the same magnitude [23], we can't conclude that genic deletions experience negative selection. However, when coupled with their low average MAF (**Table 2**) and under-abundance relative to their mutational target size (**Fig. 3A**), it seems likely that genic deletion alleles at 20% allele frequency acquire large absolute values of $\bar{\gamma}$ primarily through negative purifying selection.

**Identification of Putatively Positively Selected Indel Alleles**

Despite an expectation of neutral or negative purifying selection on large indels most of the time, a new indel allele may on occasion become the target of positive selection. Several recent studies have linked adaptive traits to novel TE insertion events [25,26]. To identify potential targets of positive selection from among the set of indels we discovered, we first polarized all indels using our outgroup accessions and then extracted a 10 kb segment centered on the indel locus for each ingroup accession. We then applied Tajima's D and Normalized Fay and Wu's H [27] to these sequences. Four derived insertion alleles – ranging from 2,140 to 5,701 bp – showed strongly negative values for both tests, which is an indication of recent positive selection (**Fig. 4**). In addition, all four insertions alleles are found at ≥ 80% frequency in the focal Iron Mountain population, also suggestive of positive selection. Three of the insertion alleles are near the upstream region of a gene (2,056, 3,981, and 13,701 bp from the start codon), while the fourth allele is 3,211 bases downstream of the nearest gene's stop codon. None of these insertions are clearly annotated as a single TE, though 2 intersect with more than one TE. The associated genes are annotated as a F-box gene (mgv1a018496m), a thaumatin family gene (mgv1a019837m), an eIF-3 family gene (mgv1a012800m), and a mitogen-activated protein kinase (mgv1a003728m). It is possible that the signatures of positive selection detected are actually associated with other polymorphisms in or near these genes. Alternatively, by analogy to the expression changes in the maize *tb1* gene driven by a large upstream insertion thought to have been selected under domestication [26], these indels may themselves be driving expression that are under positive selection.

**DISCUSSION**

Several recent studies have revealed a wealth large polymorphic indels at a species-wide level [10,14,28]. Here we extend these findings to a single population of *M. guttatus* originating from an alpine population on Iron Mountain, Oregon. These findings complement the relatively small number of other population-level studies that have been done to date [17,18,19,20], and to our knowledge mark the first study of this type in plants.

In total we find 4,142 distinct indel events segregating among ten inbred accessions extracted from Iron Mountain. On average, each accession bears 1,422.6 deletions totaling 6.6 Mb relative to the reference genome  In contrast, each accession differs on average from the reference genome at only 0.42 million SNPs. Thus, there are 16 times as many nucleotides affected by indels as by SNP in the average accession.

Assuming large indel polymorphisms in this population are near mutation-selection balance, estimates of population genetic metrics can give us insight into the evolutionary forces involved. We find that indel size and position vis-à-vis genes are both strong predictors of allele frequency (**Fig. 1** and **Table 2**), with larger indels and genic indels tending to have rarer minor alleles. We also find that 71.8% of the minor alleles are derived in the Iron Mountain population (**Table 2**), a value that is significantly higher than what was found for all classes of SNPs. Finally, we estimate that the average coefficient of selection for genic indel alleles is approximately 2-fold stronger than selection on nonsynonymous mutations and nearly 3-fold stronger than synonymous mutations when conditioning on alleles at a 20% frequency (the rarest class of alleles from which we can calculate $π_A$). These results are consistent with previous studies on the population genetics of large indels in *Drosophila* [10,19,20].

Within this pool of variation, we also find a small number of young deletion alleles that may be under positive selection. By combining two tests for positive selection, we identify four indels as outliers (**Fig. 4**), three of which involve novel high-frequency derived insertion alleles near the 5' start site of a gene. Without additional experimental evidence, we cannot yet say whether these indels affect the regulation of the adjacent genes, have another direct effect on fitness, or are not themselves targets of selection but subject to hitchhiking from selection on linked sites. Nevertheless, the approach we outline here highlights a practical genomic scan that could be used to identify candidate regulatory polymorphisms that are visible to selection.

In terms of the genes affected, the most dramatic finding is that 71 of the 598 genes segregating for a polymorphic indel belong to the NBS-LRR family, a value that is greater than twice the random expectation based on the NBS-LRR gene family size (Fisher's Exact Test $P = 1.5 \times 10^{-10}$). Interestingly, NBS-LRR genes have also been found to be enriched for large indels in soybean and *Arabidopsis* [29,30], suggesting that they are common targets for indel mutations among plants. There are several possible explanations. The NBS-LRRs gene family is thought to have an unusually high rate of gene copy turnover [31,32]. This creates many highly similar paralogs, which in turn increases the possibility that some NBS-LRR members may serve redundant functions, and therefore their loss may be selectively neutral. Similarly, Gos and Wright [33] have suggested that NBS-LRRs genes can be functionally neutral in the absence of the pathogen they detect, and under these circumstances their loss or pseudogenization would be selectively neutral with respect to selection. In addition to these neutral explanations, research has also shown that the maintenance of some NBS-LRRs can come at a cost. Bomblies *et al.* [34] found that some NBS-LRR genes play a role in hybrid necrosis, which results when a NBS-LRR incorrectly sets off plant defense pathways in a hybrid background due to off-target stimulation. This form of hybrid mortality could pose a significant fitness cost in a highly outcrossing species like *M. guttatus* and could select for non-functional indel alleles among offending NBS-LRRs during times of population admixture. Also, rather than simply being functionally neutral, at least one NBS-LRRs has been shown to decrease fitness in the absence of the pathogen it recognizes [35]. Any of these scenarios could lead to balancing selection among functional and a nonfunctional NBS-LRR alleles, and may in part explain why NBS-LRRs are highly enriched for polymorphic indel mutations.

Methods for identifying structural variants are improving rapidly, and we anticipate the methods used here may become obsolete as techniques for long-read sequencing mature. After applying a number of quality-control filters to supplement evidence from paired-end alignments alone, we achieved a modest 72% validation rate for our *in silico* predicted indels. While we did find thousands of high-confidence indel polymorphisms, we do not know how many true indels were missed by our methods. Furthermore, similar to previous efforts [15], we were unable to generate a high-confidence list of other structural variants. That said, several of the quality control filters we describe here could fruitfully be employed to help discover structural variants more generally.

## MATERIALS AND METHODS

**Plant Materials, DNA Extraction, and *Mimulus guttatus* Reference Genome Resources**

The plant materials used in this study are documented in **Table 1**. Our focal population includes 10 inbred lines extracted from a natural population on Iron Mountain, Oregon, USA. Nine were chosen at random from a pool of approximately 200 inbred lines, while the tenth, IM62, was chosen because it was also used to create the reference genome. In addition to this focal population we chose two inbred lines extracted from distant populations (DUN and SF5). All plants were grown in the Duke University Biology Greenhouse. Leaf and bud tissues were harvested for DNA extraction when plants began to flower. DNA was extracted following a urea extraction protocol modified from Shure *et al.* [36]. All accessions used in this study were inbred through self-fertilization and single seed descent at least six generations prior to DNA extraction and sequencing.

We used the *Mimulus guttatus* version 2.0 reference genome, which is available online at (ftp://ftp.jgi-psf.org/pub/compgen/phytozome/v9.0/early_release/Mguttatus_v2.0/), as are associated gene and TE annotations (ftp://ftp.jgi-psf.org/pub/compgen/phytozome/v9.0/Mguttatus/annotation/).

**Sequencing**

DNA samples were sent to the DOE Joint Genome Institute, Duke University Sequence Facility, and the University of North Carolina High-Throughput Sequencing Facility (see **Table 1**), for library preparation with the Illumina Paired-end Sample Prep. Kit V1, followed by Illumina GAII sequencing. Sequence output is available in **Table 1**, as are the NCBI-SRA accession numbers for all raw sequence data. Paired-end sequencing was performed in 2 x 35, 2 x 75, or 2 x 76 bp configurations. The mean distance between paired-end reads for all libraries was 275.8 bp and a mean within library standard deviation was +/- 28.8 bp.

**Alignment to Reference Genome and Identification of SNPs and Abnormally Aligned Read Pairs**

Sequences were aligned to the *M. guttatus* reference genome using the BWA alignment program (version 0.5.8c; [37]), with all settings left at defaults, and utilizing the paired-end read alignment option (*sampe*).

**Identifying SNPs**

SNPs were determined using the *pileup* function in the samtools package (version 0.1.8; [38]). First we extracted all read pairs with BWA mapping quality ≥ 29 and then identified sites with at least 3X coverage but no greater than 25X coverage, except for the DUN accession, which we allowed a maximum coverage of 40X. From these sites, we made a base call if > 75% of the reads displayed the same nucleotide. Finally, we only called SNPs in sites that passed these criteria for all 10 Iron Mountain accessions. All SNPs in coding regions were assigned as synonymous or nonsynonymous using the coding frame of the longest predicted transcript at that locus in the *M. guttatus* Phytozome v9.0 genome annotation. All SNP calls can be found in the Supplemental SNP data set at the Dryad digital repository [21].

**Identifying Abnormally Aligned Read Pairs**

Using samtools (version 0.1.8; [38]) to traverse the alignments, we identified all read pairs for which both members align to the *M. guttatus* reference genome with a mapping quality ≥ 29, but have abnormal relative alignment positions (pairs not in the expected orientation (→←) and/or an insert size ≥ 1000 bp). This information was assessed using information encoded in the bitwise SAM file flag values (**Table S3**). Among all lines we identified 527,059 read pairs with abnormal alignments, and these were retained for further examination.

Next, as a means of minimizing alignment errors, all abnormally aligned read pairs were realigned to the reference genome with novoalign (version 2.07.11; http://www.novocraft.com) using a k-mer size of 14 ($k$=14) and step size of 1 ($s$=1), with all other parameters left at default settings. We chose novoalign, a hash based aligner, because it uses a fundamentally different alignment algorithm than BWA, a Burrows-Wheeler transform based aligner. Novoalign identified novel high-quality (mapping quality ≥ 29) alignments that were not abnormal for 871 read pairs. After removing these read pairs we were left with 526,188 abnormally aligned read pairs that had been confirmed by both BWA and novoalign.

**Clustering Abnormally Aligned Reads to Identify Putative SVs**

Following the strategy of Chen *et al.* [8], abnormally aligned reads from all accessions were pooled to make use of all available information when predicting SVs and localizing their breakpoints. After pooling, abnormally aligned read pairs were clustered into sets that came from the same genomic locations for both the forward and reverse read pairs. This clustering was done using the ClusterTree function in the bx-python package (version 0.7.0; http://pypi.python.org/pypi/bx-python), which provides a data structure for finding clusters of intervals where both endpoints fall within a certain window size. Based on the smallest mean insert size among our paired-end sequences, we chose 225 bp as our maximum window size. Furthermore, we also required that the putative SV clusters be supported by at least three read pairs, regardless of which accessions contributed those read pairs. All retained clusters were partitioned into SV classes (deletions, inversions, transpositions) based on the paired read configuration. Finally, the accession(s) contributing to each cluster were assigned. Python code for read pair clustering is available at the Dryad digital repository [21].

**Filtering SVs**

We resequenced the accession (IM62) that was used to construct the *M. guttatus* reference genome. By aligning paired-end reads from IM62 to itself, we were able to identify spurious SV calls and remove any SVs that included IM62 as one of the accessions containing the putative event. Also, we only retained SVs ≥ 1000 bp. Because we were working with highly inbred lines, we expect nearly all loci to be homozygous, and accessions containing a predicted deletion relative to the IM62 reference genome should show few or no high-quality read alignments to the deleted interval. Following this principle, we only retained deletion events in which the putatively deleted interval had a read depth of coverage in the lowest 10th percentile of the genome-wide distribution for that accession. Also, we found that some regions in the *M. guttatus* genome produce a large number of abnormally aligned reads (**Fig. S1**). We suspect that repeats make these intervals difficult for read alignment, so we removed any SVs from these regions. This was done by counting the number of abnormally aligned reads for both endpoints of a candidate SV using a 5,000 bp window with the focal SV in the middle. After discounting all abnormally aligned reads assigned to the focal SV, we determined if the count of additional abnormally aligned reads in this window was greater than the 90th percentile of all 5,000 bp windows in the genome. If it was, the focal SV was dropped. Also, for inversions and transpositions, we enforced that read coverage across the SV interval remain between the 10th and 90th percentile of the genome-wide distribution, to avoid regions that have either unusually sparse or dense coverage. Finally, if one accession failed the test when applying the filters listed above, the candidate SV was rejected for all accessions. Python code for filtering SVs is available from the Dryad digital repository [21]. The cluster assignment, SV type, and filtering fate for all 527,059 abnormally aligned read pairs is available as a supplemental data set from the Dryad digital repository [21].

**Monte Carlo Methods, Coalescent Simulations, and Population Genetic Calculations**

To obtain an estimate of the expected average absolute allele frequency difference for a sample size of 10 individuals relative to the "true" allele frequency estimate from approximately 100 individuals we performed a Monte Carlo simulation. We first drew a true allele frequency randomly from a uniform distribution between 0 and 1. Then randomly drew 10 samples from a binomial distribution of size 1 using the true allele frequency as the probability of success. This sample makes up the simulated observed allele frequency from a sample size of 10, matching our sample size from the Iron Mountain population. Because our indel detection scheme was such that the minor allele had to be found in at least 1 accession, we censored the simulated observed values so they fell between 0.1 and 0.9. Finally we collected the absolute difference between the true value and the censored observed value. This process was repeated 50,000 times, and the average of these replicates was 0.097

The expected frequency spectrum for neutral loci was estimated using the ms coalescent simulation software [39]. We simulated 10,000 genealogies, each with 10 unique haploid chromosomes (the effective number of genomes sampled from 10 inbred accessions) and 100 segregating sites with no recombination. From these replicate genealogies we calculated the average site frequency spectrum, the expected minor allele frequency (MAF) and the upper and lower bounds of the 95% confidence interval for Tajima's D (-1.72, +1.59), and Normalized Fay and Wu's H (-1.88, +5.70). Tajima's D and Normalized Fay and Wu's H [27] were calculated using the EggLib package (version 2.1.5; [40]). Normalized Fay and Wu's H requires specification of an ingroup and an outgroup. This was achieved by randomly selecting a value between one and five, and randomly assigning this many simulated chromosomes to the outgroup.

Intra-allelic nucleotide diversity ($\pi_A$) was calculated for all synonymous and nonsynonymous

SNPs and for genic deletion alleles. We focused on only the genic deletion alleles, as we expect that many of the insertion alleles are wild-type because they were predicted to maintain an open reading frame in the IM62 genome annotation. For each polymorphism type, we separated the two alleles and calculated $\pi_A$ for each allele class on a 500 bp haplotype centered on the polymorphism. Singleton alleles (i.e. 10% frequency) were ignored because their $\pi_A$ cannot be defined. $\pi_A$ values were then aggregated by allele frequency and mean $\pi_A$ at each frequency was used an estimate of allelic age. Diffusion equations used to transform $\pi_A$ into estimates of the mean population scaled coefficient of selection (    ) are given in Kiezun *et al.* [24]

**Availability of data, software and materials**

Seeds from the inbred lines used in this study are available from the *Mimulus* Stock Center (www.mimulusevolution.org/stocks.php). Sequencing data have been deposited at the NCBI-SRA (**Table 1**). Additional data and software are available at the Dryad digital repository [21].

**ACKNOWLEDGMENTS**

LEF was supported by a NIH Ruth L. Kirchstein NRSA award [F32GM090763] as well by NSF EF-0328636 to JHW and TJV, and NIH R01-GM078991 to TJV. The authors wish to thank the many collaborators in the Joint Genome Institute Community Sequencing Program's *Mimulus* project for work on the reference genome, gene and TE annotations, and for resequencing data from both outgroups used in this study. We thank Por Tangwancharoen and Melina Smith for help with the validation study, and Uffe Hellsten, Benjamin Blackman and Thomas Clarke for helpful comments on the structural variation and SNP prediction pipelines. We also thank the Duke Genome Sequencing and Analysis Core, the UNC Chapel Hill High-Throughput Sequencing Facility, and the Joint Genome Institute for producing the Illumina sequences used in this study. Finally, we thank the UNC Chapel Hill Research Computing Center for access to their high-performance compute cluster.

## AUTHOR CONTRIBUTIONS

Conceived and designed the experiments: LEF JHW TJV. Performed the experiments: LEF. Analyzed the data: LEF. Contributed reagents/materials: JHW TJV. LEF and TJV wrote the manuscript and all authors approve of the final version.

**REFERENCES**

1. Britten RJ (2002) Divergence between samples of chimpanzee and human DNA sequences is 5%, counting indels. Proc Natl Acad Sci USA 99: 13633-13635.

2. The Chimpanzee Sequencing and Analysis Consortium (2005) Initial sequence of the chimpanzee genome and comparison with the human genome. Nature 437: 69-87.

3. Hu TT, Pattyn P, Bakker EG, Cao J, Cheng J-F, et al. (2011) The *Arabidopsis lyrata* genome sequence and the basis of rapid genome size change. Nat Genet 43: 476-481.

4. Sturtevant AH, Dobzhansky T (1936) Inversions in the third chromosome of wild races of *Drosophila pseudoobscura*, and their use in the study of the history of the species. Proc Natl Acad Sci USA 22: 448-450.

5. Feuk L, Carson AR, Scherer SW (2006) Structural variation in the human genome. Nat Rev Genet 7: 85-97.

6. Borevitz JO, Liang D, Plouffe D, Chang H-S, Zhu T, et al. (2003) Large-scale identification of single-feature polymorphisms in complex genomes. Genome Res 13: 513-523.

7. Handsaker RE, Korn JM, Nemesh J, McCarroll SA (2011) Discovery and genotyping of genome structural polymorphism by sequencing on a population scale. Nat Genet 43: 269-276.

8. Chen K, Wallis JW, McLellan MD, Larson DE, Kalicki JM, et al. (2009) BreakDancer: an algoithm for high-resolution mapping of genomic structural variation. Nat Meth 6: 677-681.

9. Chia J-M, Song C, Bradbury PJ, Costich D, de Leon N, et al. (2012) Maize HapMap2 identifies extant variation from a genome in flux. Nat Genet 44: 803-807.

10. Emerson JJ, Cardoso-Moreira M, Borevitz JO, Long M (2008) Natural selection shapes genome-wide patterns of copy-number polymorphism in *Drosophila melanogaster*. Science 320: 1629-1631.

11. Kidd JM, Cooper GM, Donahue WF, Hayden HS, Sampas N, et al. (2008) Mapping and sequencing of structural variation from eight human genomes. Nature 453: 56-64.

12. Korbel JO, Urban AE, Affourtit JP, Godwin B, Grubert F, et al. (2007) Paired-end mapping reveals extensive structural variation in the human genome. Science 318: 420-426.

13. Swanson-Wagner RA, Eichten SR, Kumari S, Tiffin P, Stein JC, et al. (2010) Pervasive gene content variation and copy number variation in maize and its undomesticated progenitor. Genome Res 20: 1689-1699.

14. Cao J, Schneeberger K, Ossowski S, Gunther T, Bender S, et al. (2011) Whole-genome sequencing of multiple *Arabidopsis thaliana* populations. Nat Genet 43: 956-963.

15. 1000 Genomes Project Consortium (2012) An integrated map of genetic variation from 1,092 human genomes. Nature 491: 56-65.

16. Redon R, Ishikawa S, Fitch KR, Feuk L, Perry GH, et al. (2006) Global variation in copy number in the human genome. Nature 444: 444-454.

17. Montgomery SB, Goode DL, Kvikstad E, Albers CA, Zhang ZD, et al. (2013) The origin, evolution, and functional impact of short insertion–deletion variants identified in 179 human genomes. Genome Res 23: 749-761.

18. Feulner PGD, Chain FJJ, Panchal M, Eizaguirre C, Kalbe M, et al. (2013) Genome-wide patterns of standing genetic variation in a marine population of three-spined sticklebacks. Mol Ecol 22: 635-649.

19. Cridland JM, Thornton KR (2010) Validation of rearrangement break points identified by paired-end sequencing in natural populations of *Drosophila melanogaster*. Genome Biol Evol 2: 83-101.

20. Mackay TFC, Richards S, Stone EA, Barbadilla A, Ayroles JF, et al. (2012) The *Drosophila melanogaster* genetic reference panel. Nature 482: 173-178.

21. Dryad Repository http://doi.org/dryad.###. [To be added]

22. MacArthur DG, Balasubramanian S, Frankish A, Huang N, Morris J, et al. (2012) A systematic survey of loss-of-function variants in human protein-coding genes. Science 335: 823-828.

23. Maruyama T (1974) The age of a rare mutant gene in a large population. Am J Hum Genet 26: 669-673.

24. Kiezun A, Pulit SL, Francioli LC, van Dijk F, Swertz M, et al. (2013) Deleterious alleles in the human genome are on average younger than neutral alleles of the same frequency. PLoS Genet 9: e1003301.

25. Aminetzach YT, Macpherson JM, Petrov DA (2005) Pesticide resistance via transposition-mediated adaptive gene truncation in *Drosophila*. Science 309: 764-767.

26. Studer A, Zhao Q, Ross-Ibarra J, Doebley J (2011) Identification of a functional transposon insertion in the maize domestication gene *tb1*. Nat Genet 43: 1160-1165.

27. Zeng K, Fu Y-X, Shi S, Wu C-I (2006) Statistical tests for detecting positive selection by utilizing high-frequency variants. Genetics 174: 1431-1439.

28. Conrad DF, Pinto D, Redon R, Feuk L, Gokcumen O, et al. (2010) Origins and functional impact of copy number variation in the human genome. Nature 464: 704-712.

29. McHale LK, Haun WJ, Xu WW, Bhaskar PB, Anderson JE, et al. (2012) Structural variants in the soybean genome localize to clusters of biotic stress-response genes. Plant Physiol 159: 1295-1308.

30. Shen J, Araki H, Chen L, Chen J-Q, Tian D (2006) Unique evolutionary mechanism in R-genes under the presence/absence polymorphism in *Arabidopsis thaliana*. Genetics 172: 1243-1250.

31. Meyers BC, Kaushik S, Nandety RS (2005) Evolving disease resistance genes. Curr Opin Plant Biol 8: 129-134.

32. Baumgarten A, Cannon S, Spangler R, May G (2003) Genome-level evolution of resistance genes in *Arabidopsis thaliana*. Genetics 165: 309-319.

33. Gos G, Wright SI (2008) Conditional neutrality at two adjacent NBS-LRR disease resistance loci in natural populations of *Arabidopsis lyrata*. Mol Ecol 17: 4953-4962.

34. Bomblies K, Lempe J, Epple P, Warthmann N, Lanz C, et al. (2007) Autoimmune response as a mechanism for a Dobzhansky-Muller-type incompatibility syndrome in plants. PLoS Biol 5: e236.

35. Tian D, Traw MB, Chen JQ, Kreitman M, Bergelson J (2003) Fitness costs of R-gene-mediated resistance in *Arabidopsis thaliana*. Nature 423: 74-77.

36. Shure M, Wessler S, Fedoroff N (1983) Molecular identification and isolation of the Waxy locus in *maize*. Cell 35: 225-233.

37. Li H, Durbin R (2010) Fast and accurate long read alignment with Burrows-Wheeler transform. Bioinformatics.

38. Li H, Handsaker B, Wysoker A, Fennell T, Ruan J, et al. (2009) The sequence alignment/map format and SAMtools. Bioinformatics 25: 2078-2079.

39. Hudson RR (2002) Generating samples under a Wright-Fisher neutral model of genetic variation. Bioinformatics 18: 337-338.

40. De Mita S, Siol M (2012) EggLib: processing, analysis and simulation tools for population genomics. BMC Genetics 13: 27.

41. Fishman L, Saunders A (2008) Centromere-associated female meiotic drive entails male fitness costs in monkeyflowers. Science 322: 1559-1562.

**Table 1: Resequenced accessions, including outgroups.**

| Line | Species | Origin[a] | Generations of Inbreeding | Seq. Facility[b] | Total paired end reads | Read type | Sites available[c] | Median per site coverage[c] | NCBI SRA Accession # |
|---|---|---|---|---|---|---|---|---|---|
| IM109 | M. guttatus | Iron Mountain | 11 | UNC | 24,671,221 | 2 X 75 bp | 127,863,968 | 8 | SRX021073 |
| IM1145 | M. guttatus | Iron Mountain | 11 | UNC | 22,839,207 | 2 X 75 bp | 138,331,815 | 8 | SRX021074 |
| IM155 | M. guttatus | Iron Mountain | 12 | Duke | 37,172,361 | 2 X 75 bp | 138,971,514 | 15 | SRX055301 |
| IM320 | M. guttatus | Iron Mountain | 15 | Duke | 23,226,015 | 2 X 75 bp | 160,689,132 | 8 | SRX055300 |
| IM479 | M. guttatus | Iron Mountain | 9 | UNC | 24,086,031 | 2 X 75 bp | 134,867,795 | 9 | SRX021077 |
| IM62 | M. guttatus | Iron Mountain | >10 | UNC | 24,911,877 | 2 X 75 bp | 206,733,050 | 7 | SRX021072 |
| IM624 | M. guttatus | Iron Mountain | 13 | UNC | 22,433,144 | 2 X 75 bp | 137,605,733 | 8 | SRX021075 |
| IM693 | M. guttatus | Iron Mountain | 9 | UNC | 21,969,210 | 2 X 75 bp | 133,128,765 | 8 | SRX021078 |
| IM767 | M. guttatus | Iron Mountain | 11 | UNC | 25,497,466 | 2 X 75 bp | 135,649,759 | 9 | SRX021079 |
| IM835 | M. guttatus | Iron Mountain | 13 | UNC | 17,966,309 | 2 X 75 bp | 129,433,596 | 6 | SRX021076 |
| DUN | M. guttatus | Florence | >6 | JGI | 262,093,335 | 2 X 35 bp | 94,024,553 | 23 | SRX030973,SRX030974 |
| SF5 | M. nasutus | Sherar's Falls | natural selfer | JGI | 24,199,117 | 2 X 76 bp | 65,812,268 | 10 | SRX116529 |

[a] All accessions originate from Oregon, USA. Approximate geographic coordinates as follows: Iron Mountain [44.4005, -122.1428], Florence [43.8891, -124.1360], and Sherar's Falls [45.2587, -121.0201], with [Latitude, Longitude] given in decimal format.

[b] Duke University Sequence Facility (Duke); DOE Joint Genome Institute (JGI); University of North Carolina High-Throughput Sequencing Facility (UNC)

[c] Nucleotide sites belonging to reads with mapping quality scores ≥ 29

**Table 2: Indel and SNP variants among the ten resequenced lines.**

| Polymorphism Type | Count | Median size (bp) | Average MAF[a] | Proportion of derived minor alleles |
| --- | --- | --- | --- | --- |
| indels – all | 4,142 | 2,563 | 0.255 | 0.718 |
| Indels in genes | 414 | 3,804 | 0.218 | 0.739 |
| Indels in TEs | 1,855 | 2,839 | 0.277 | 0.743 |
| SNPs - all | 1,337,759 | 1 | 0.222 | 0.619 |
| SNPs – synonymous | 259,676 | 1 | 0.239 | 0.550 |
| SNPs – nonsynonymous | 106,638 | 1 | 0.227 | 0.626 |

[a] average MAF from neutral coalescent simulations = 0.222

**FIGURES**

**Figure 1: Distribution of indel size as a function of allele frequency.** The boxplots indicate the distribution of indel sizes (bp) at different minor allele frequencies (MAF) for all indels identified in the focal Iron Mountain population. Indel size (*y*-axis) is plotted on a log scale.

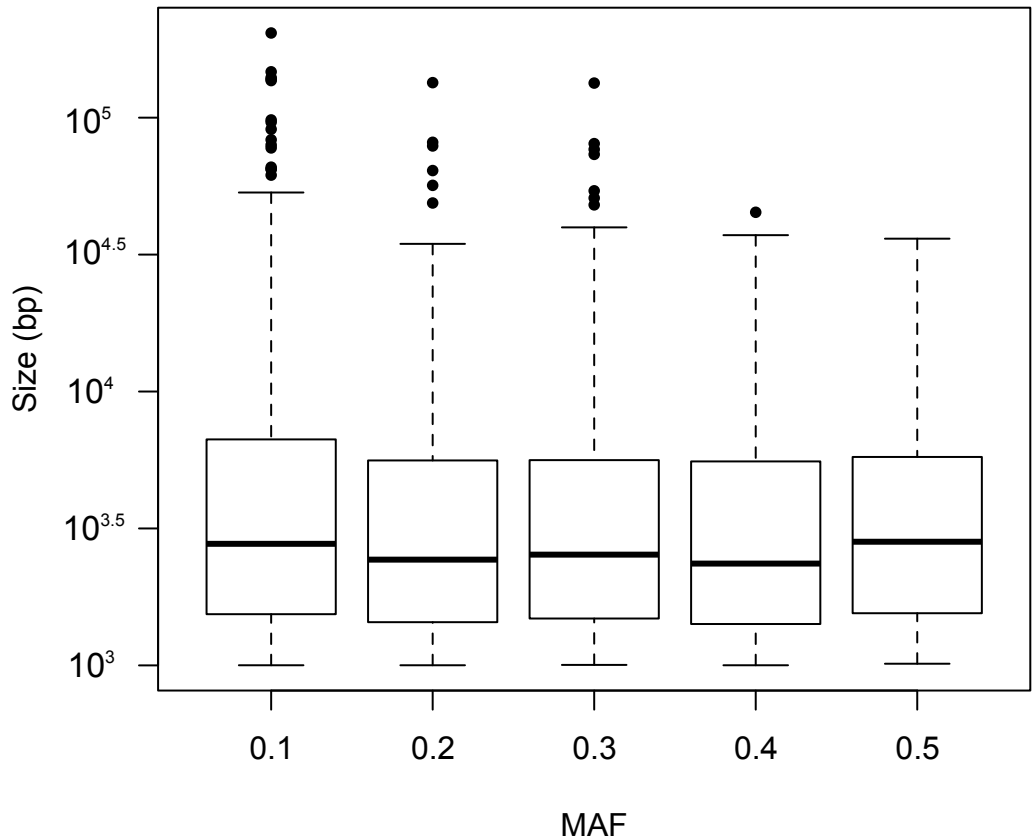

**Figure 2: Allele frequency distributions in the ten resequenced lines.** Cumulative frequency as a function of minor allele frequency is shown for genic and transposable element (TE) indels, synonymous and nonsynonymous SNPs, and a neutral coalescent simulation (neutral). Inset: Estimates of mean Tajima's D color-coded as in the main panel. Bars indicate the 95% confidence intervals as obtained by delete-one jackknifing the ten lines. Sample sizes are given in **Table 2**.

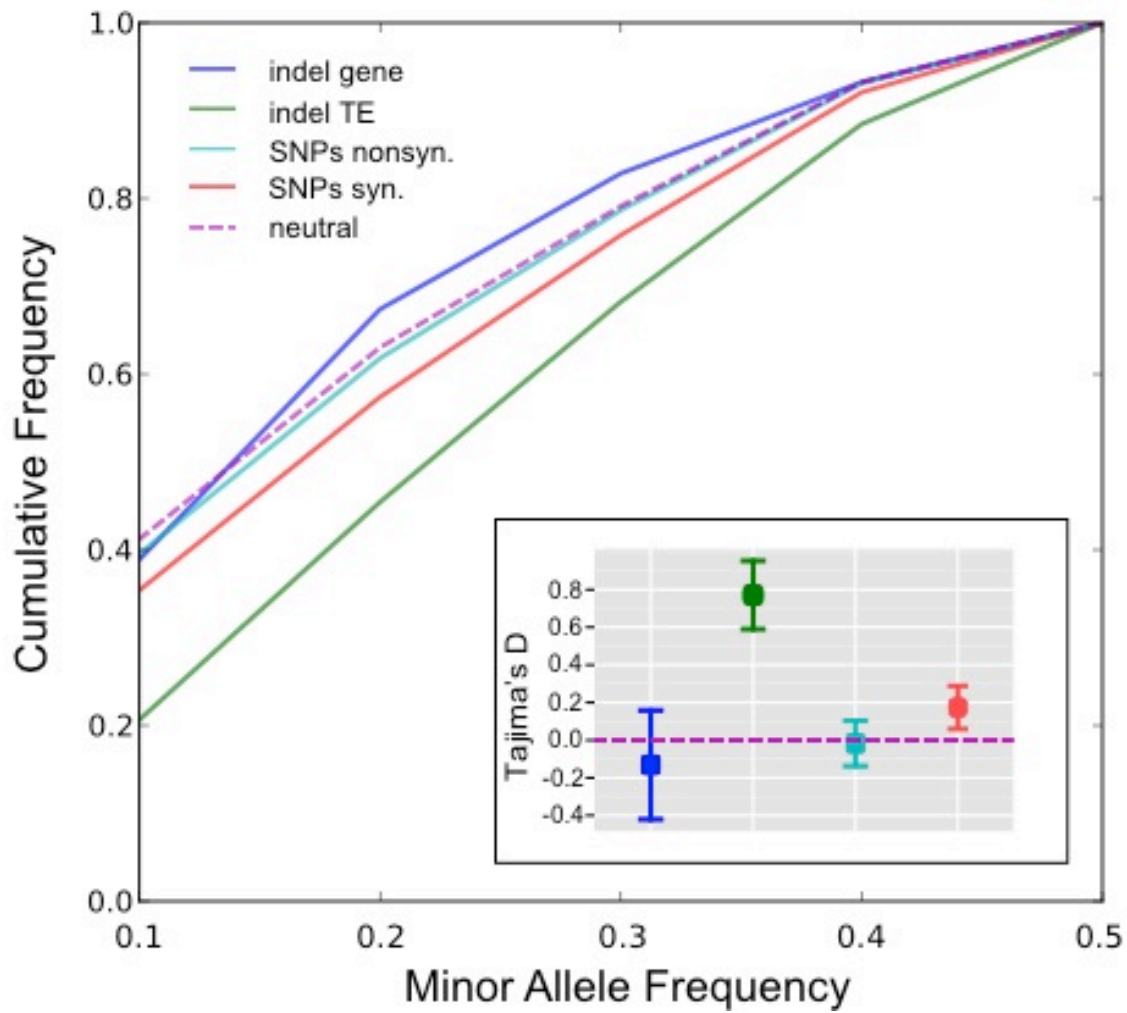

**Figure 3: Distribution of indel polymorphisms in genes and transposable elements**. The observed and expected number of nucleotide sites in segregating indels among (A) genes and transposable elements (TE), and (B) different gene components. (C) Indel density along a normalized transcript, using data from all annotated genes overlapping with segregating deletions. Each genic region (5' UTR, gene body consisting of exons plus introns, and 3' UTR) was divided into 100 equally sized bins, and for each bin the relative density among all polymorphic indels was recorded (*y*-axis). The distribution of bin densities is expected to approximately follow a binomial distribution. The red dashed lines indicate the upper and lower 95% confidence bounds for a binomial distribution with *p*=0.01 and *n* given by the total number of inserted/deleted base pairs observed in that region.

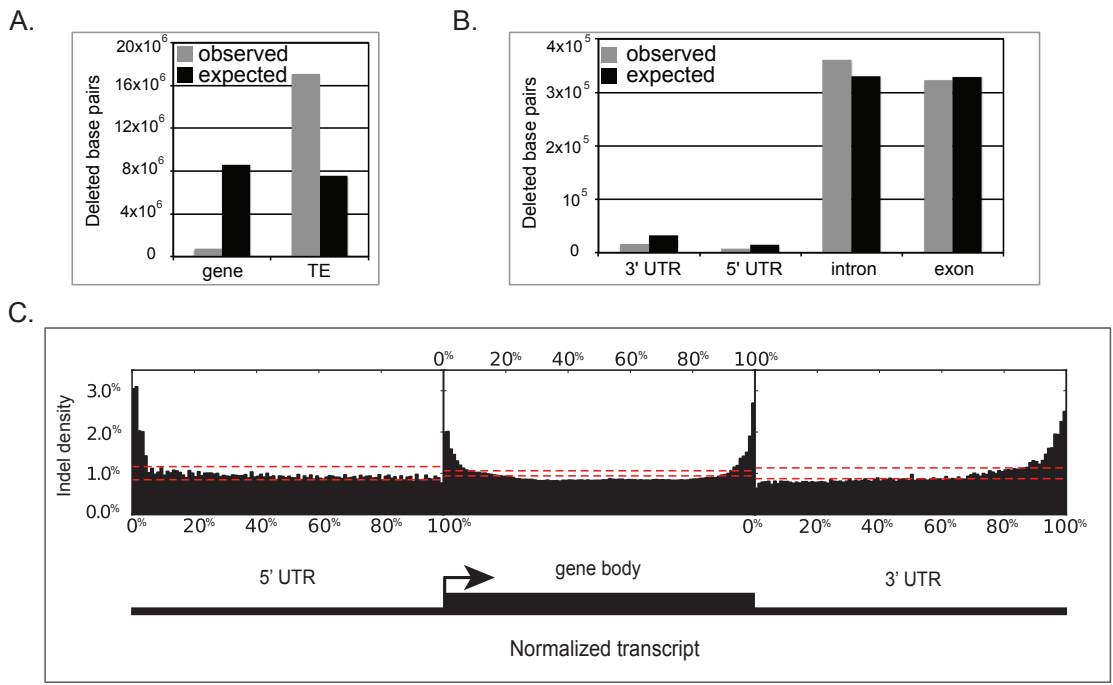

**Figure 4: Identification of indel loci putatively under positive selection.** The plot includes Normalized Fay and Wu's H and Tajima's D estimates for each indel. The red box in the lower left corner indicates the area outside the lower 95% confidence interval of both metrics as assessed by coalescent simulation.

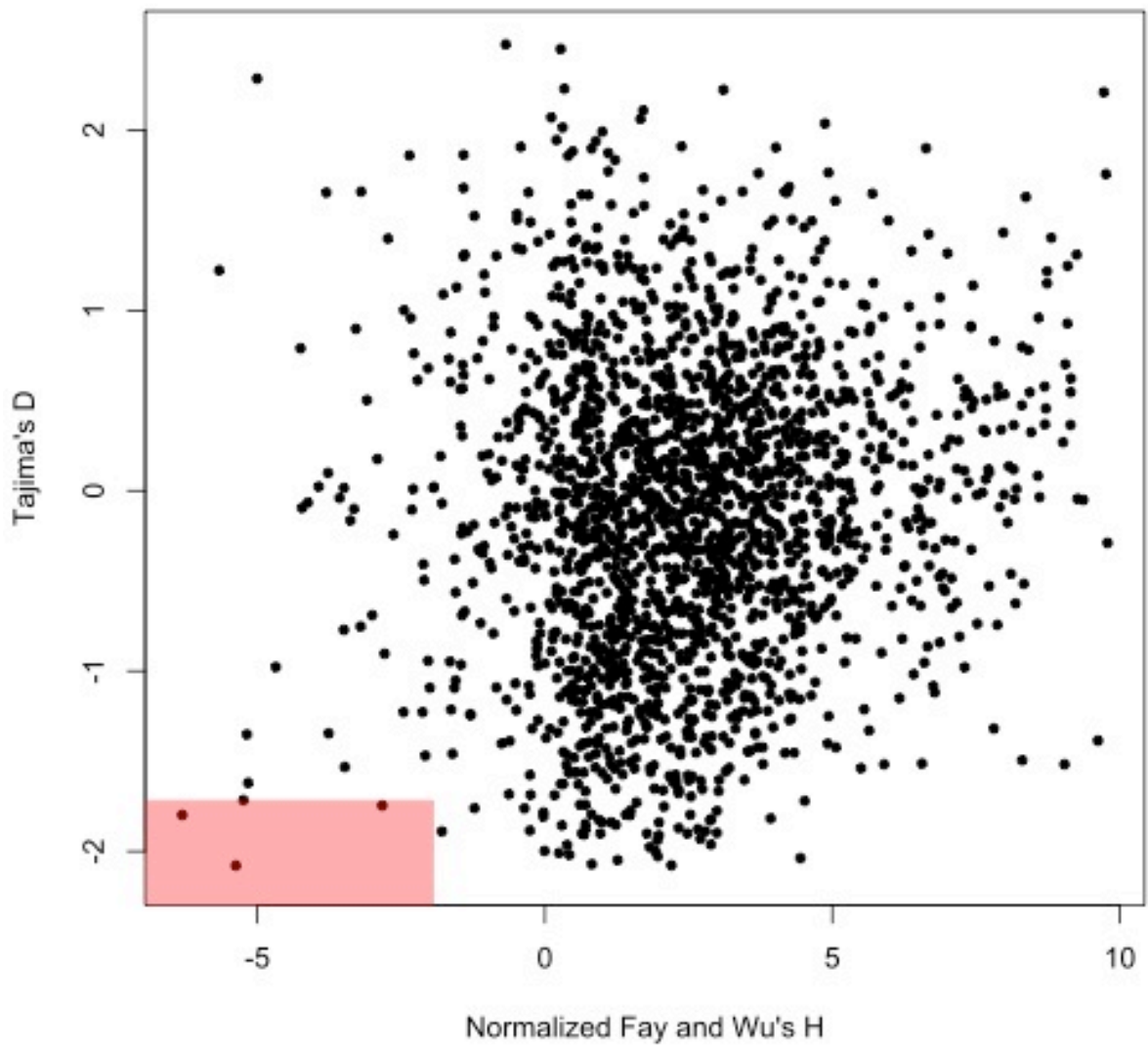

**SUPPORTING INFORMATION**

**Figure S1: Alignment of read pairs along *Mimulus* chromosome 9.** The two dot-plots show the left and right read pair positions for high-quality paired read alignments (BWA mapping quality ≥ 29) in the reference line (IM62; left) and a non-reference line (IM693; right). Paired reads that align to the reference genome in the expected location fall along the diagonal, while all abnormally aligned reads are represented above the diagonal. Concentrations of abnormal read pairs in IM62 indicate regions in which read pair data from other lines would be unreliable for detection of SVs. The abundance of abnormally aligned read pairs indicates the extent of the technical challenge that must be overcome to filter the signal from the noise.

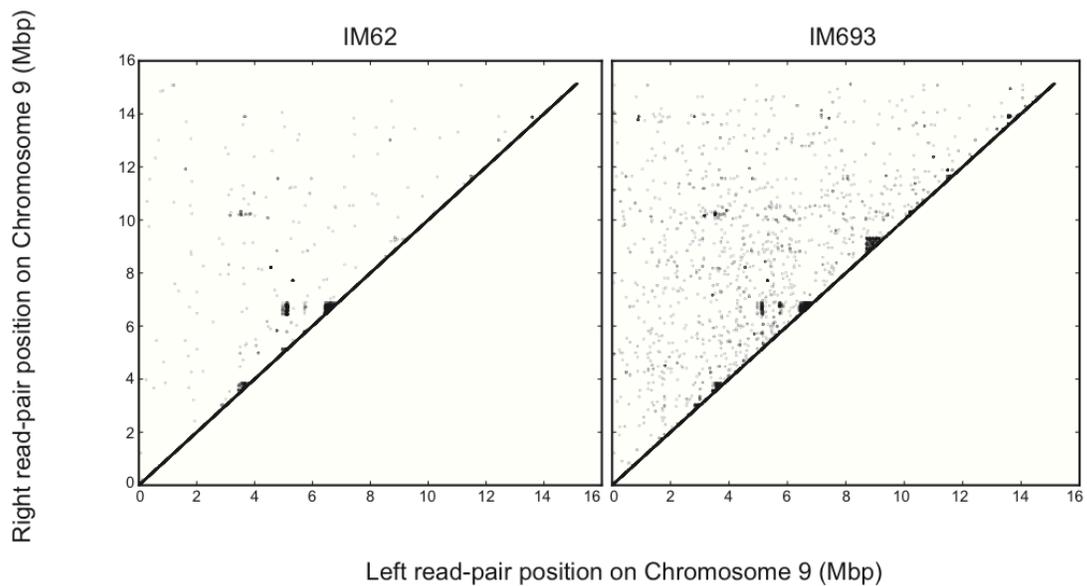

**Figure S2: Nucleotide diversity (π) within putative inversion intervals for both the inverted and the reference genome form of an inversion.** The black line shows the pattern expected if nucleotide diversity within putative inversion haplotype does not differ between inversion haplotypes. The seven inversion events (red points) showing unusually low nucleotide diversity among inverted haplotypes were our targets for validation. We chose these with the rationale that the rare form of the inversion might be recent and have little nucleotide diversity, and that this signature would be unlikely to come from collinear regions, thus enriching these seven for *bona fide* inversions. However, we could not generate unique primer pairs for any of these seven.

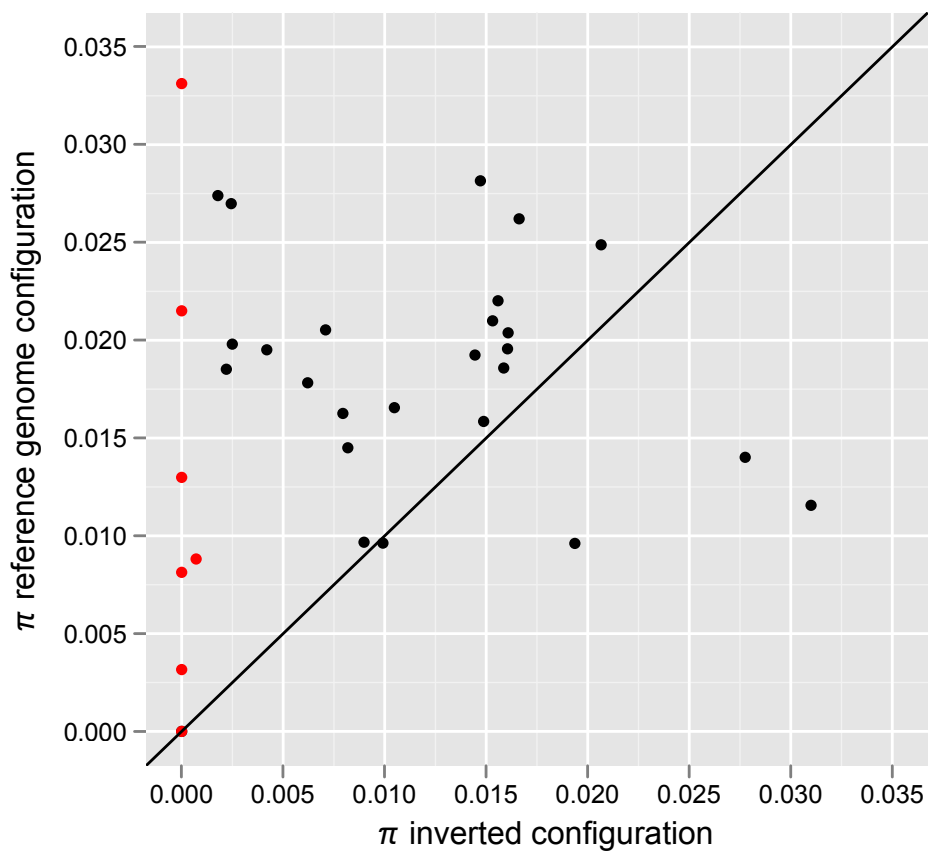

**Figure S3: Deletion allele frequency estimates from deep population sampling (*x*-axis) and whole genome resequencing of ten inbred lines (*y*-axis).** The linear regression is shown in red with the corresponding $R^2$.

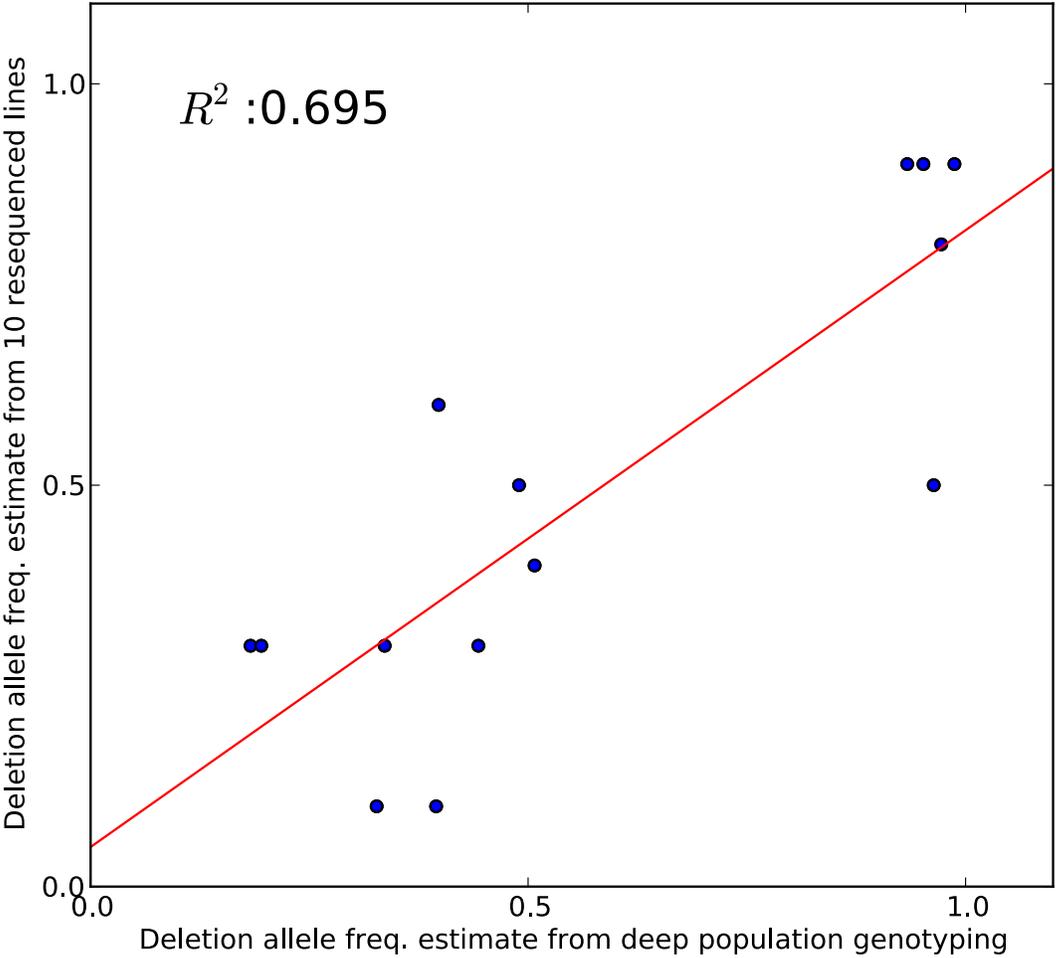

**Figure S4: Relationship of deletion frequency to coverage and nucleotide divergence.**

Left: there is only a weak negative relationship between the number of read pairs aligned and the number of deletions found within each of the nine non-reference lines. Right: there is a positive relationship between the pairwise nucleotide divergence and the number of deletions found between IM62 and each of the nine non-reference lines. Linear regressions are shown in red with corresponding $R^2$ values.

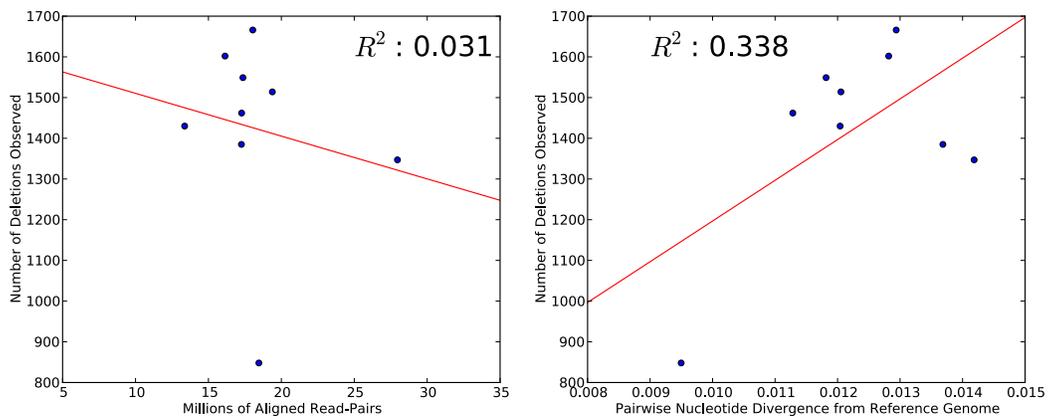

**Figure S5: Distribution of genes, repeats, centromeres, alignable regions, and indels along *M. guttatus* chromosome 1.** Chromosome 1 has been broken into 25 kb segments, and for each segment various proportions or counts are represented by an associated legend to the right of each subplot. Chromosome 1 appears to be acrocentric, with genes clustered primarily on the left side and the centromeric/repetitive fraction dominating the right side of the. The "Proportion Alignable" refers to the proportion of sites in a 25 kb segment that support at least one aligned paired-end read with a mapping quality scores ≥ 29. As expected, the gene rich left side of the chromosome is enriched for alignable sites, and as a consequence is expected to be more available for indel discovery. The other 13 *M. guttatus* chromosomes (not shown) have similar patterns. Centromeric repeats were annotated by blast search using the *M. guttatus* 728 bp centromeric repeat sequence [41].

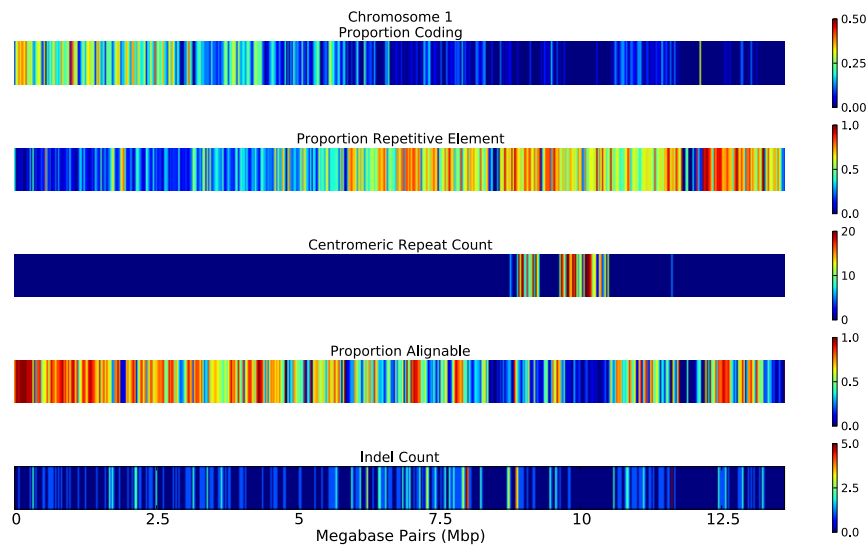

**Figure S6: Estimating the coefficient of selection**

(A) Table of all intra-allelic nucleotide diversity results from 500 bp haplotypes surrounding genic deletions, nonsynonymous and synonymous mutations for alleles at 20% to 90% frequency. (B) Theoretical relationship between strength of selection and mean allelic age for alleles at 20% frequency.

A) Table of all intra-allelic $\pi_A$ results from 500 bp haplotypes surrounding genic deletions, nonsyn. and syn. mutations.

| Allele Frequency | Polymorphism | N | Pi | sd | stderr |
|---|---|---|---|---|---|
| 0.2 | genic del | 96 | 0.003147 | 0.006612 | 0.000675 |
| 0.2 | nonsyn | 25060 | 0.004736 | 0.010214 | 6.45E-05 |
| 0.2 | syn | 57304 | 0.005532 | 0.010907 | 4.56E-05 |
| 0.3 | genic del | 54 | 0.007954 | 0.012098 | 0.001646 |
| 0.3 | nonsyn | 18971 | 0.005462 | 0.01083 | 7.86E-05 |
| 0.3 | syn | 47964 | 0.005718 | 0.010826 | 4.94E-05 |
| 0.4 | genic del | 30 | 0.00723 | 0.011152 | 0.002036 |
| 0.4 | nonsyn | 16156 | 0.006746 | 0.011587 | 9.12E-05 |
| 0.4 | syn | 42325 | 0.006875 | 0.011572 | 5.62E-05 |
| 0.5 | genic del | 28 | 0.007062 | 0.010963 | 0.002072 |
| 0.5 | nonsyn | 7754 | 0.0082 | 0.012585 | 0.000143 |
| 0.5 | syn | 20626 | 0.0083 | 0.012605 | 8.78E-05 |
| 0.6 | genic del | 12 | 0.010982 | 0.011816 | 0.003411 |
| 0.6 | nonsyn | 16155 | 0.009061 | 0.013021 | 0.000102 |
| 0.6 | syn | 42328 | 0.009155 | 0.012951 | 6.29E-05 |
| 0.7 | genic del | 8 | 0.002711 | 0.004547 | 0.001608 |
| 0.7 | nonsyn | 18972 | 0.010001 | 0.013526 | 9.82E-05 |
| 0.7 | syn | 47969 | 0.009814 | 0.013229 | 6.04E-05 |
| 0.8 | genic del | 15 | 0.007961 | 0.009405 | 0.002428 |
| 0.8 | nonsyn | 25068 | 0.010581 | 0.013678 | 8.64E-05 |
| 0.8 | syn | 57314 | 0.010565 | 0.013512 | 5.64E-05 |
| 0.9 | genic del | 10 | 0.007436 | 0.006816 | 0.002156 |
| 0.9 | nonsyn | 44401 | 0.011933 | 0.014409 | 6.84E-05 |
| 0.9 | syn | 92141 | 0.011294 | 0.013862 | 4.57E-05 |

N = Sample size
Pi = intra-allelic $\pi_A$
sd = standard deviation
stderr = standard error

B) **Relationship between strength of selection and mean allelic age for alleles at 20% frequency***

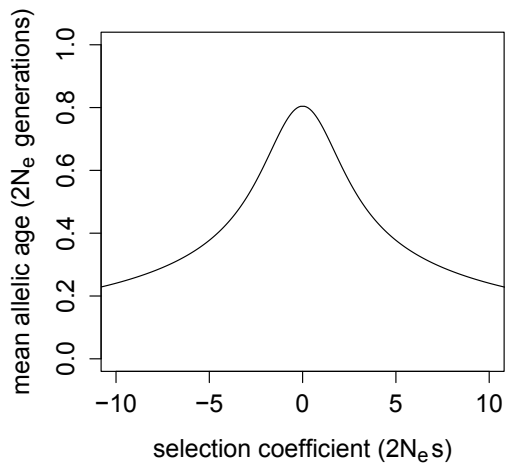

*following Kiezun et al., 2012, PLoS Genet., 9:e1003301

**Table S1: Validation results.** The table includes the genomic location of each targeted indel, the primer pairs used for PCR and validation results.

[Not included]

**Table S2: Indels associated with transposable element families.** The first worksheet reports the transposable element results aggregated to the level of families of related elements, while the second includes all results individualized to specific elements sub-families found within *M. guttatus.* For each TE counts are given for the 2 × 2 contingency table used to contrast observed polymorphisms versus expected on the basis of genomic abundance. In addition results from the Fisher's Exact Test, the odds ratio, and observed fold change are given.

[Not included]

**Table S3: Observed proportions of various configurations of bitwise SAM flags.** For each pair of bitwise SAM flags the putative structural interpretation is given along with its frequency among the all reads for each resequenced line.

[Not included]